\documentstyle [12pt]{article}
\newcommand{\be}{\begin{equation}}
\newcommand{\ee}{\end{equation}}
\newcommand{\ra}{\rightarrow}

\newcommand{\lsim}{\stackrel{<}{\sim}}
\newcommand{\mpl}{M_{\rm Pl}}
\begin{document}
\begin{titlepage}
\begin{flushright}
 IFUP-TH 44/93\\
 October 1993\\
 hep-th/9310157
\end{flushright}

\vspace{8mm}

\begin{center}

{\Large\bf  Black Hole Complementarity and

\vspace{4mm}

 the Physical Origin of the Stretched Horizon}\\

\vspace{12mm}
{\large Michele Maggiore}

\vspace{3mm}

I.N.F.N. and Dipartimento di Fisica dell'Universit\`{a},\\
piazza Torricelli 2, I-56100 Pisa, Italy.\\
\end{center}

\vspace{4mm}

\begin{quote}
\hspace*{7mm} ABSTRACT. We discuss
the idea of black hole complementarity, recently suggested by Susskind
{\em et al.}, and the notion of stretched horizon, in
the light of the generalized uncertainty principle of
quantum gravity. We discuss implications for the no-hair theorem and
we show that within this approach quantum hair arises naturally.
\end{quote}
PACS categories: 04.60, 12.25, 97.60.
\end{titlepage}
\clearpage

The problems related  to the application of quantum
mechanics to black holes rank between the
most challenging in theoretical physics. Despite great
effort, it has not yet been  reached a consensus on the
validity of Hawking's
claim~\cite{Haw} that the evolution of states in the
presence of black holes  violates unitarity.
Recently, an extremely interesting proposal,
close in spirit to previous work of 't~Hooft~\cite{tH},
has been put forward
by Susskind and coworkers [3-6] and termed ``black hole complementarity''.
The basic observation is
that physics looks very different to an observer in free fall in a
black hole and to a ``fiducial observer'' at rest with respect to the
black hole, outside the horizon.
Crossing the horizon of a very massive black hole,
the free falling observer  should not experience anything out of the
ordinary. If the mass of the hole $M$ is much larger than
Planck mass $\mpl$ a classical description of the black hole should be
adequate, and in classical general relativity the horizon merely
represents a coordinate singularity, while
physical quantities like the curvature are non-singular. Furthermore,
from the point of view of the free falling observer,
the flux of Hawking radiation is switched off  when he
approaches the horizon.
This can be shown observing
that near the horizon, with an appropriate change of variables, the
Schwarzschild metric approaches the Rindler metric, and a free falling
observer in Schwarzschild spacetime
becomes a free falling observer in flat Minkowski space --
and certainly does not detect any radiation. The point of view of a
fiducial observer is dramatically different. In Schwarzschild
coordinates, a fiducial observer at a distance $r$ from
a Schwarzschild  black hole
measures an effective  temperature
\be
T=(1-\frac{2GM}{r})^{-1/2} \, T_H\, ,
\ee
where $T_H=\hbar/(8\pi GM)$ is Hawking temperature.  Climbing out of the
gravitational potential well, the radiation is gravitationally
red-shifted by a factor $(1-\frac{2GM}{r})^{1/2}$ and is seen by an
observer at infinity as having temperature $T_H$.
Instead, at the horizon $r=2GM$
the  temperature measured by a fiducial observer diverges. For a
fiducial observer, this temperature is certainly a very real effect.
If too close to the horizon, he would  be killed by the eccessive
heat. From this point of view, a fiducial observer regards the black
hole horizon as a physical membrane, endowed with real physical
properties. More in general, within the
membrane paradigm~\cite{TPM} all interactions of a
black hole with the external environment, as seen by a fiducial
observer, are described in terms of a two-dimensional membrane
 endowed with properties like electric
conductivity, viscosity, entropy and temperature.

The difference between the point of view of  free falling and
fiducial observers can be
 of relevance to the information loss problem.
There are various approaches to this problem (for  reviews see
e.g.~\cite{HS}), and each one has its own difficulties; in particular,
if one assumes that the black hole evaporates completely then
the core of the information loss problem
is that it appears very difficult to reconcile the point of view of
the free falling and fiducial observers, without questioning basic
postulates of quantum mechanics. In fact, as discussed in~\cite{Sus1},
the assumption that the evolution of states is unitary,
togheter with
the superposition principle,  forces upon us the conclusion that
all distinctions between the infalling states must be obliterated soon after
they  cross the horizon; this  is certainly very difficult to
reconcile with the point of view of the free falling observer and with
the equivalence principle, since to the  free falling observer
the horizon is no special place.
Of course, after passing the horizon
 the free falling observer  cannot
communicate anymore with  fiducial observers, so that no immediate
logical contradiction arises; for instance, the free falling observer
cannot report on the lack of substance of the membrane.
More in general, the investigation
of various gedanken experiments carried out in~\cite{Sus3}
indicates that {\em ``apparent logical contradictions can always be
traced to unsubstantiated assumptions about physics at or beyond the
Planck scale''} \cite{Sus3,Pre}.  This
implies that, contrarily to the common  opinion, the information
loss paradox cannot be addressed without a detailed knowledge of
a full quantum theory of  gravity.

Both in the membrane paradigm and in discussing black hole
complementarity a key role is played by the concept of
stretched horizon. It has been found~\cite{TPM} that the description
of the black hole in terms of a membrane takes a much simpler
and elegant form  if the horizon is stretched, i.e. if the surface
of the black hole is moved at a slightly larger radius, and a set of
membrane-like conditions are imposed at the stretched horizon. This
allows to get rid of many irrelevant details of the infalling fields
and at the same time ``regularizes''  the infinities coming from the
infinite red-shift factor between $r=2GM$ and $r=\infty$.\footnote{A
conceptually similar approach is given by the ``brick wall'' model of
't~Hooft~\cite{tH}.}
In particular,
the temperature measured by a fiducial observer at the stretched
horizon is large but finite. The amount of stretching is however
rather arbitrary. Because of this, and because the free falling
observer would not agree on its existence,
the membrane has been considered as
a useful mathematical construction rather than a physical object: for
instance (see~\cite{TPM}, pag. 31) ``...it is very useful to regard
these boundary conditions as arising from physical properties of a
{\em fictitious} membrane residing
at the location of the stretched horizon.
More specifically, it is useful to {\em pretend} that
the stretched horizon
is endowed with a surface density of electric charge ...'' (our
italics).
However, performing a gedanken experiment aimed at measuring the
radius of the horizon with the best possible accuracy, we have
recently found~\cite{MM1}, using only rather general
arguments,  that the error $\Delta x$ on the radius of
the horizon is subject to a generalized uncertainty principle,
\be\label{1}
\Delta x\ge \frac{\hbar}{\Delta p} +{\rm const.}\, G\Delta p\,
\ee
which implies the existence of a minimum error on the order of the
Planck length times a numerical constant, which is shown in~\cite{MM1}
to be larger than one.
It is tempting to assume that eq.~(\ref{1}) actually represents a
generalized uncertainty principle which governs all measurements in
quantum gravity; a similar uncertainty principle
 has been found in string theory [11-14]. Independently
of the correctness of the latter  assumption, eq.~(\ref{1})
holds for the measurement of a black hole radius, which is the case in
which we are now interested. More exactly, eq.~(\ref{1}) only holds
for $\Delta p$ not  large compared with $\mpl$, since  $\Delta p$ is
the error on the momentum of a particle emitted by the black hole and
detected at infinity, and we do not really know how to describe the
particle if its energy is super-planckian.
The two terms on the
right-hand side can be considered as the first  terms in an
expansion in powers of $\Delta p/\mpl$. The knowledge of the exact
expression would in principle require a full quantum theory of
gravity.
In the following we will assume that the exact expression
valid for arbitrarily large values
of $\Delta p/\mpl$  does not spoil the main result which can be
inferred from eq.~(\ref{1}), namely the fact that there exists a
minimum error on the horizon radius.
We can also try to guess  the exact form of the generalized
uncertainty principle making the assumption that
it can be derived from an algebraic structure, in the same
sense in which the standard uncertainty principle is a consequence of
the Heisenberg algebra.  In~\cite{MM3} we have found that there
is indeed an appropriate algebraic structure, and it is given by a
deformation of the Heisenberg algebra involving a deformation
parameter $\kappa$ with dimensions of mass,
\begin{eqnarray}
\left[ x_i ,x_j \right] &= & -\frac{\hbar^2}{\kappa^2}\,
i\epsilon_{ijk}J_k\label{xx}\\
\left[ x_i , p_j \right]   &= & i\hbar\delta_{ij}
(1+\frac{E^2}{\kappa^2})^{1/2}\, .\label{xp}
\end{eqnarray}
($E$ is the energy and $J_i$ the angular momentum). In the limit
$\kappa\ra\infty$ it reduces to the Heisenberg algebra. In the
following $\kappa$ will be identified with  $\mpl$, apart from
numerical factors (alternatively, we can identify $\hbar/\kappa$ with
the string length times a numerical factor of order one, in order to
recover the string uncertainty principle). It is remarkable that,
under relatively mild assumptions, this deformed
algebra is unique, essentially because the Jacobi identities provide
very stringent requirements on the possible deformations of an algebra.

{}From eq.~(\ref{xp}) we immediately
derive the generalized uncertainty principle
\be\label{gu}
 \Delta x_i\Delta p_j \ge  \frac{\hbar}{2} \delta_{ij}
\langle\left( 1+\frac{E^2}{\kappa^2}\right)^{1/2}
\rangle\, .
\ee
Expanding the square root in powers of
$(E/\kappa )^2$ and using
$\langle {\bf p}^2\rangle ={\bf p}^2+(\Delta p)^2$, where
$(\Delta p)^2=\langle ({\bf p}-\langle {\bf p}\rangle )^2\rangle$,
at first order one obtains
\be
 \Delta x_i\Delta p_j \ge  \frac{\hbar}{2} \delta_{ij}
\left( 1+\frac{E^2+(\Delta p)^2}{2\kappa^2}\right)
\, .
\ee
which reproduces eq.~(\ref{1}) in the limit
$E\ll\kappa ,\Delta p\lsim\kappa$.
Instead, in the limit
$\langle{\bf p}\rangle^2\sim (\Delta p)^2\gg\kappa^2$ one obtains
\be
\Delta x\ge {\rm const}\times \frac{\hbar}{\kappa}\, .
\ee
These results suggest that the  horizon is subject to
irreducible quantum fluctuations which provides it with a physical
thickness. In this case, we can
attempt to promote the membrane from a useful
mathematical construction to a real physical entity:
in spite of the fact that the nature of its microphysical degrees of
freedom is at present quite elusive, still from the
point of view of  fiducial observers the membrane has definite
and real physical properties, and it has a physical thickness and a
location in space which are well defined and determined by physics,
rather than by our ``regularization'' procedure.
In particular, we see that it extends
beyond the nominal horizon by a few Planck lengths (if we choose
$\kappa\sim\mpl$). This agrees with the choice suggested in~\cite{Sus1}
in the case of two-dimensional dilaton gravity, while in~\cite{TPM}
the stretched horizon is assumed  to extend outward
by  a finite fraction of the Schwarzschild radius. It is important
to observe that the thickness of the membrane is independent of the
black hole mass; this implies that even the horizon
of a ``classical'' black hole,
with $M\gg\mpl$, acquires a thickness because of quantum effects.

Of course, the membrane  does not exists for the free falling
observer; however, the principle of black hole complementarity
protects us from logical inconsistencies. To ask whether the membrane
exists or not is like asking whether a photon went through
a specific arm
of an interferometer. The answer depends on the
setting of the experiment. In our case, on whether the observer is
in free fall or not.

Promoting the membrane to a real, physical object implies a radical
revision of some of the common wisdom concerning black holes. In
particular, one  realizes that there is no reason to expect that the
classical
no-hair theorem extends in the quantum domain as well. Let us remind
the form of the classical no-hair theorem for
the simple case of a massive scalar field
(see~\cite{CPW}
for a   discussion of the relevance of the
no-hair theorem to the information loss problem and to the possible
relation between black holes and elementary particles). Introducing
the tortoise coordinate
\be
r_*=r+2GM\log\frac{r-2GM}{2GM}\,
\ee
the wave equation for a  scalar field of mass $\mu$
in the Schwarzschild
background, after expanding in partial waves, reads
\be\label{wave}
\left( -\frac{\partial^2}{\partial t^2}
+ \frac{\partial^2}{\partial r_*^2}\right) \psi_{l,m}=
\left( 1-\frac{2GM}{r}\right)
\left( \mu^2+\frac{2GM}{r^3}+\frac{l(l+1)}{r^2}\right) \psi_{l,m}
\, .
\ee
In the zero-frequency limit the second derivatives with respect to
$r_*$  is always positive. However, the domain $2GM\leq r<\infty$
corresponds to $-\infty <r_*<\infty$, and therefore a solution which
decreases exponentially at $r=\infty$ must blow up at the horizon.
Thus there is no physically acceptable static solution: ``a black
hole has no hair''. In the
quantum case, of course the right-hand side of eq.~(\ref{wave})
receives corrections, which we cannot control.
However, it is usually argued that these corrections should not alter
the asymptotic behavior at  spatial infinity, nor close to the
horizon, where the dominant effect is given by the factor
$1-2GM/r$. Then, the no-hair theorem simply follows from the fact that
$r_*$ ranges from $-\infty$ to $+\infty$.

{}From the membrane point of view, it is not difficult to see where this
argument can fail. Physically, we are not allowed to extrapolate the
solution inside the membrane.  Such an extrapolation would imply to
enter a region of super-planckian temperatures and therefore to make
assumptions about physics beyond the Planck scale -- which is just
what the principle of black hole complementarity warns us not to do.
As long as we stop at the border of the membrane,
as determined physically by the generalized uncertainty principle,
$r_*$ only covers a
semi-infinite range, and  a solution which decays exponentially at
spatial infinity is finite on the stretched horizon. Thus,
there is no reason to expect that
the no-hair theorem goes through even in quantum gravity.

The fact that the membrane paradigm makes possible
to  violate  the no-hair
theorem at the quantum level is not at all surprising. After all, in
the approach of refs. [3-6] the membrane is just the place where the
infalling information is stored, before being  re-radiated in such a
way as to preserve quantum coherence, according to the mechanism
suggested by Page~\cite{Page}. The
various states of the membrane correspond to different internal states
of the black hole, and the difference between the internal states
manifest itself to an observer at infinity through differences in
the Hawking radiation; this is nothing but  quantum  hair.

Another point which deserves attention is that,
at least as far as the radius of the black hole
horizon is concerned, there exists a minimal {\em spatial} distance
on the order of the Planck length. This
has the surprising consequence  (already
pointed out  in this context by Susskind~\cite{Sus4})
that at this length
scale Lorentz transformations must saturate. This implies a deep
revision of kinematics at the Planck scale. A possible example of a
different kinematic framework is provided by quantum deformations of
the Poincar\'e algebra~\cite{LNR}.
In~\cite{MM2} we  found in fact that,
in the $\kappa$-deformed Poincar\'e algebra, the $\kappa$-deformed
Newton-Wigner position operator and the
generators of translations and rotations
actually obey the algebra (\ref{xx},\ref{xp}). Another possible
kinematic framework is provided by string theory, see below.

The principle of black hole complementarity has also important
consequences for the mental image that we have of black holes. The
important lesson
that we learn is that we should be very careful not to mix
up the point of view of free falling observers with that
of fiducial observers.
Much of the seemingly paradoxical features of the information loss
problem come from a confusion between these two  points of view
and is rooted in the implicit
and seemingly undisputable assumption that
there exists a notion of invariant event.
However, Susskind~\cite{Sus4} has made the crucial observation that
black hole complementarity implies that even
the notion of invariant event cannot be anymore relied upon.  He
further observes that string theory has just the properties required
by black hole complementarity, as far as
the notion of event is concerned:
if a string falls toward a black hole, an observer at infinity
sees the string spreading when it reaches the stretched horizon, until
it covers the horizon completely, while a free falling observer sees a
string with constant transverse and longitudinal size which crosses
the horizon without any peculiar behavior. We wish to point out that
the non-invariance of the concept of event in string theory can be
seen also at a more fundamental level and is in fact well-known
(see~\cite{GSW}, pag. 29).
In general, events are defined
in terms of interactions: in classical physics the collision between
two billiard balls constitues a typical
 event. In quantum field theory, a typical event is the
emission of a photon by a source. If we represent
it by a Feynman diagram, the vertex of the interaction defines the
spacetime location of the
event. We can describe this location in different reference frames,
but the location itself has an invariant meaning.
In string theory, we must instead  consider the splitting of
strings as defining  events. However,  the point
in spacetime at
which a string splits into  two strings
appears different to different observers (see fig.~1.6 of ref.~\cite{GSW})
and correspondingly there is no Lorentz invariant notion of event.
Thus, the non-invariance of the notion of event is not specific to
physics in the vicinity of black holes, although
black holes act as a sort of magnifier of this effect.

\end{document}